\documentclass[aps,twocolumn,nofootinbib,preprintnumbers]{revtex4-1}
\usepackage{bm}% bold math
\usepackage[hidelinks]{hyperref}% Auto-generate hyperlinks without boxes
\usepackage{amsmath,amssymb,amsfonts,graphicx}
\renewcommand{\vec}[1]{{\mathbf{#1}}}
\newcommand{\beq}{\begin{eqnarray}}
\newcommand{\eeq}{\end{eqnarray}}
\usepackage{amsmath}
\usepackage{amssymb}
\usepackage{caption}
\usepackage{subcaption}
\usepackage[paper=letterpaper,margin=1in]{geometry}
\usepackage{braket}
\usepackage{upgreek}

\renewcommand{\vec}[1]{\boldsymbol{#1}}

\usepackage{tikz}

\newcommand{\tabi}{\hspace{.1\textwidth}}

\begin{document}

\title{Anomalous Dimension of the Electrical Current in the Normal State of the Cuprates from
 the Fractional Aharonov-Bohm Effect}
\author{Kridsanaphong Limtragool and Philip W. Phillips}\thanks{Guggenheim Fellow}

\affiliation{Department of Physics and Institute for Condensed Matter Theory,
University of Illinois
1110 W. Green Street, Urbana, IL 61801, U.S.A.}

\date{\today}

\begin{abstract}
We show here that if the current in the normal state of the cuprates has an anomalous dimension, then the Aharonov-Bohm flux through a ring does not have the standard $eBA/\hbar$ form, where $A$ is the area, $B$ is the external magnetic field, and $e$ is the electric charge, but instead it is modified by a geometrical factor that depends directly on the anomalous dimension of the current. We calculate the Aharonov-Bohm flux in square and disk geometries. In both cases, the deviation from the standard result is striking and offers a fingerprint about what precisely is strange about the strange metal.
\end{abstract}
\maketitle

\section{Introduction}

Before Faraday discovered that moving charges induce magnetic fields ($\vec B$), electric and magnetic fields were thought to be independent.  A concise mathematical synthesis of the two requires an additional entity, the vector potential, $\vec A$, which in classical physics is experimentally undetectable. Aharonov and Bohm\cite{AB} showed, however, that in quantum mechanics, the principle of gauge invariance imbues the vector potential with physical content such that the wave function of a charged particle moving in a closed loop around a magnetic solenoid experiences a phase shift that is determined entirely by the line integral,
\begin{equation}
\Delta\phi=\frac{e}{\hbar}\oint \vec A\cdot d \vec \ell,
\end{equation}
of the vector potential around a closed loop. Because                                                                                                                                                                                                                                                                                                                                                                                                                                                                                                                                                                                                                                                                                                                                                                                                                                                                                                                                                                                                                                                                                                                                                                                                                                                                                                                                                                                                                                                                                                                                                                                                                                                                                                                                                                                                                                                                                                                                                                                                                                                         
$\vec\nabla\times\vec A=\vec B$
and Stokes' theorem which allows us to convert a line integral to a surface one, the integral simplifies to $eBA/\hbar$, where $A$ is the cross sectional area of the magnetic solenoid, $e$ is the electric charge, and $\hbar$ is Planck's constant divided by $2\pi$. The key physical surprise here is that charges outside the solenoid know about the magnetic field solely because of the spatial extent of the vector potential.
The relationship between the vector potential and the magnetic and electric fields implies that all the equations of classical electromagnetism are invariant with respect to the transformation, 
\beq
\label{eq:U1transformation}
 A_\mu\rightarrow A_\mu+\partial_\mu\Lambda
\eeq
where $\partial_\mu=(-\partial_t/c,\partial_x,\partial_y,\partial_z)$ and $\Lambda$ is an arbitrary dimensionless function. Because $\Lambda$ is dimensionless, this transformation fixes the dimension of $A_\mu$ to be unity; that is, $A_\mu$ has dimensions of inverse length.  A further consequence of the invariance of electricity and magnetism to a choice in the gauge is that there has to be a corresponding conserved current whose dimension is set by the generators of the $U(1)$ symmetry group.  The resulting dimension of the conserved current in a relativistic theory is $d$ where $d$ is the spatial dimension.   Clearly if $[A]\ne 1$,  the underlying theory is not governed by the standard 1-form gauge-invariant principle of electricity and magnetism. 

There are no known examples in nature of a conserved current in which the vector potential has a dimension other than unity. Perhaps possible exceptions to this rule could obtain in exotic materials such as the high-temperature copper-oxide superconductors. This problem remains unsolved because no knock-down experiment has revealed unambiguously the nature of the charge carriers in the normal state. What we know for sure is that the standard theory of metals and a single-parameter\cite{pchamon,karchhartnoll} formulation of quantum criticality cannot simultaneously explain $T$-linear resistivity, power-law optical conductivity\cite{Marel2003,Hwang2007,Basov2011}, breakdown of the Weidemann-Franz law\cite{lorentz}, and the scaling of the Hall angle\cite{hallangle}.  
However, recent theoretical work\cite{limtragool,karchanom,karchhartnoll}  suggests that all of the transport properties of the normal state can be explained by positing a conserved current with an anomalous dimension.

Indeed, this is a striking prediction because a textbook problem\cite{gross,peskin,wen1992scaling} in quantum field theory is to prove that conserved quantities cannot acquire anomalous dimensions under renormalization.   For a local theory away from the strict relativistic limit, the dimension of the current can change by two mechanisms: 1) reduction of the effective dimensionality, that is a violation of hyperscaling\cite{Gouteraux2011,Huijse2012} with exponent $\theta$ or 2) space and time scale differently thereby requiring a dynamical exponent $z>1$\cite{Kachru2008}. The new scaling of the current is now $d-\theta+z-1$.  Either of these can be modeled using holography with a bulk dilaton construction\cite{g1,g2}.  Of course other scenarios exist in which the $U(1)$ symmetry is explicitly broken by the presence of a mass for the gauge-field\cite{Gouteraux:2012yr,Gouteraux2011}.  However, the lack of a conserved current in this case makes this scenario quite distinct from the hyperscaling violation variants\cite{Huijse2012}.  A third approach for the emergence of an anomalous dimension for the current is that the underlying theory is inherently non-local.  It is this mechanism that appears to be operative in the recent work\cite{karchanom,limtragool}  which showed that extending the single-parameter quantum critical scenario\cite{karchhartnoll} to include a multi-band or unparticle sector with a running charge\cite{Karch:2015pha} leads to a consistent explanation of all the power laws experimentally observed in the dc\cite{hallangle,lorentz,qcrit2,qcrit3} and ac\footnote{In the cuprates, the power law in the ac conductivity\cite{Marel2003,Hwang2007,Basov2011} appears in the mid-infrared and hence does not persist down to zero frequency. For the multi-band construction with a mass-dependent relaxation time\cite{limtragool,karchanom} to match this feature, the summation over mass needs to have a cutoff\cite{limtragool}. In this case, there exists an onset energy scale $\tau_0^{-1}$ for the power law to appear.}\cite{Marel2003,Hwang2007,Basov2011} transport properties in the strange metal phase of the cuprates. A running charge is possible only if the vector potential acquires an anomalous dimension, $\Phi$\cite{Gouteraux:2012yr,g1,g2}. To fit the cuprates $\Phi=-2/3$\cite{karchanom,karchhartnoll}.  Given the novelty of an electric current acquiring an anomalous dimension as the unique underlying feature of the strange metal, it would be ideal to design an experiment, not tethered to any scaling analysis, that can critically test this idea unambiguously. Should this be borne out experimentally, then the normal state of the cuprates would represent the first example in nature of current carrying excitations with an anomalous dimension.  

In this paper, we propose such an experiment. Since it is the vector potential that communicates the anomalous dimension to the electrical current, this effect should be detectable from a simple Aharonov-Bohm\cite{AB} (AB) experiment in the strange metal regime.  We show that the new principle that maintains gauge invariance implies that the AB phase must pick up a factor that depends on the anomalous dimension and hence provides an unambiguous fingerprint of the non-locality of the current. The physical set up is a sample pierced by a constant magnetic field. The associated gauge field permeates the sample and picks up the anomalous dimension. The resultant AB phase, no longer $\Delta\phi=eBA/\hbar$, picks up extra factors of $L^{\alpha_B-2}$, where $L$ is a quantity with units of length and $\alpha_B$ is the scaling dimension of the B-field, for  $\Delta\phi$ to be dimensionless.  We calculate this effect explicitly.

\section{Fractional Gauge Transformation}

Indeed it is gauge invariance that makes the problem of anomalous dimensions for the gauge field highly problematic {\it a priori}.  Consider the transformation in Eq. \ref{eq:U1transformation} applied to the action
\beq
S=\int d^dx[F^2+J_\mu A^\mu+\cdots].
\eeq
Since the field strength, $F$ is invariant under Eq. (\ref{eq:U1transformation}), the action transform
\beq
S\rightarrow S+\int d^dx J_\mu\partial^\mu\Lambda.
\eeq
Consequently, invariance under Eq. (\ref{eq:U1transformation}), upon integration by parts, results in the standard 
charge conservation equation
\beq\label{chargecons}
\partial^\mu J_\mu=0.
\eeq

The natural question that arises is if an anomalous dimension is not compatible with Eq. (\ref{eq:U1transformation}), then what is the consequence for charge conservation?
Indeed fractional formulations of electricity and magnetism do exist\cite{lazo2011,herrmann,domokos} based on the gauge principle
${\alpha}_\alpha A_{\mu}(x) \rightarrow {_\alpha}A_{\mu}(x) + \partial^{\alpha_\mu}_{\mu}\Lambda(x)$ 
which contain fractional derivatives (see Appendices \ref{app:frac_calc_fourier} and \ref{app:frac_calc_real}) of the phase $\Lambda(x)$, the power of which fixes the engineering dimension of $_{\alpha}A_\mu$ to be $\alpha_\mu$.   From the argument presented previously, such an implementation will affect the charge conservation equation.  But an immediate problem with such constructions is that the gauge transformation is not rotationally invariant and hence this is not an acceptable theory. 

What the charge conservation equation lays plain is that any operator, $\hat Y$, which commutes with the total differential can be used to redefine the current operator and hence will change its dimension {\bf without} affecting the linear nature of Eq. (\ref{chargecons}). However, a key restriction on the operator ${\hat Y}$ is that it cannot change the order of the form of either the current or the dual current ($\star J$).  If such an operator exists, it would also offer a loophole around the general argument advanced by Gross\cite{gross} that it is the commutator of the charge density with any $U(1)$ field, $\phi(x)$,
\beq
\delta(x_0-y_0)[J_0(x),\phi(y)]=\delta \phi(y)\delta^D(x-y),
\eeq 
that fixes the scaling dimension of the conserved current.  Here $\delta \phi(y)$ is the change in the field $\phi$ to linear order upon acting with the  $U(1)$ transformation and $J_0$ is the charge density.  Letting $J_0\rightarrow {\hat Y}J_0$ we see that the current no longer has dimension $D$ but rather  $D-D_Y$ where $D_Y=[{\hat Y}]$.  

Elsewhere\cite{csforms} we have shown how to construct ${\hat Y}$ explicitly for dilaton actions of the form,  
\beq \label{eq:action_zf2}
S=\int d^{d+1}x \sqrt{-g}Z(\phi)F^2+\cdots,
\eeq
used by holographic models\cite{g1,g2} to yield either anomalous dimensions for the gauge field or hyperscaling violation exponents.  Here, $Z(\phi)\sim e^{\gamma\phi}$ is a dilaton field and $F$ the field strength.  The equations of motion for the Maxwell part of the action are
\beq
\nabla^\mu(Z(\phi) F_{\mu\nu})=0,
\eeq
where $\nabla^\mu$ is the covariant divergence.
A typical solution\cite{g1,g2} for the dilaton field is $\phi\sim \ln\kappa r$. Consequently the equations of motion are equivalent to
\beq
\nabla^\mu( y^aF_{\mu\nu})=0.
\eeq
In the language of differential forms, this equation becomes
\beq
d(y^a\star dA)=0
\eeq
which clearly illustrates that for any slice perpendicular to the radial direction, the standard $U(1)$ gauge transformation applies.
To determine what happens at the boundary, we note that
these equations are reminiscent of  those studied by Caffarelli and Silvestre\cite{CS2007} (CS)  for the case of a scalar field,
\beq
\nabla\cdot(y^a \nabla u),
\eeq
which is just a recasting of the elliptic differential equation
\beq
u(x,y=0)=f(x)\\
\Delta_x u+\frac{a}{y}u_y+u_{yy}=0.
\eeq
What they were interested in is what form does this differential equation acquire at the boundary, $y\rightarrow 0$.
What Cafarelli/Silvestre showed is that any equation of this kind  satisfies
\beq
\lim_{y\rightarrow 0}(y^a u_y)=C_{d,\gamma}(-\Delta)^\gamma f(x).
\eeq
where $\gamma=(1-a)/2$.  

The exact same result holds for the gauge field as it is just a 1-form generalization of the CS extension theorem.  
In a separate paper, we have generalized\cite{csforms} the CS extension theorem to p-forms.  The result is as expected.  The
 p-form generalization of the CS extension theorem 
yields the boundary action of the form,
\beq
\label{AEQ}
S =\frac12 \int A_i (-\nabla)^{2\gamma} A^i,
\eeq
whose propagator in Lorentz gauge, $\partial^\gamma_i A^i = 0$, is 
\beq
D^{ij}(k) = \langle A^i(k) A^j(-k)\rangle = \frac{-i\eta^{ij}}{(k^2)^\gamma}.
\eeq
Clearly at the boundary $[A_i]\ne 1$.  
 The corresponding field strength is the 2-form,
\beq
\label{fboundary}
F = d_\gamma A= d(-\Delta )^\frac{\gamma-1}{2} A,
\eeq
with gauge-invariant condition,
\beq\label{eq:u1frac}
A \rightarrow A+ d_\gamma \Lambda,
\eeq
 with 
\beq
 \ d_\gamma \equiv (-\Delta)^\frac{\gamma-1}{2}d,
 \eeq
 which preserves the 1-form nature of the gauge-field with dimension $[A_\mu]=\gamma_\mu$, rather than unity.  Note  $[d,(-\Delta)^{\frac{\gamma-1}{2}}]=0$.  Consequently,
 we identify ${\hat Y}=(-\Delta)^{\frac{\gamma-1}{2}}$ which is a completely rotationally invariant operator.  In general, the total differential commutes with any power of the Laplacian operator and hence the conservation equation is uniquely specified up to $(-\Delta)^\alpha$.  This added ambiguity in the formulation of electricity and magnetism does not seem to have been noticed until now. 
  
What the p-form generalization\cite{csforms} of the CS extension theorem lays plain in the context of holographic models that
 yield an anomalous dimension for the gauge field is that the anomalous dimension enters the boundary theory (see Eq. (\ref{eq:u1frac})) as a result of 
 the rotationally invariant entity,
\beq
\partial^\gamma_\mu \equiv (-\Delta)^\frac{\gamma-1}{2}\partial_\mu,
\eeq
which we take to be our operational definition of the fractional derivative.   As expected, the action in terms of the electromagnetic field strength defined by Eq. (\ref{fboundary})
\beq
S = \int -\frac{1}{4}F_{ij}F^{ij}
\eeq
is identical to Eq. (\ref{AEQ}). Simply integrate by parts and pick a gauge $\partial^\alpha_i A^i = 0$ and the  action reduces to the action, 
Eq. (\ref{AEQ}),  which results from the CS extension theorem.
Consequently, the boundary actions of the holographic models that generate anomalous dimensions or hyperscaling violation exponents all contain fractional Laplacians and hence transform under the non-local gauge transformation, Eq. ( \ref{eq:u1frac}).  

It is instructive to compute the current-current correlator for the action with $F=d_\gamma A$.  Consider the action with a coupling to matter field (through the current $J^i$)
\beq
S = \int -\frac{1}{4}F_{ij}F^{ij} + J^iA_i.
\eeq
The equation of motion of this action is
\beq
\partial^\gamma_i F^{ij} = J^j.
\eeq
We will identify the current by this equation. Hence the current-current correlation function we compute is valid at the level of equation of motion and hence the current will explicitly have an unorthodox dimension. The current-current correlation function is then given by
\beq
C^{ij}(x,y) &=& \langle \partial^\gamma_l F^{li}(x)\partial^\gamma_p F^{pj}(y)  \rangle \nonumber \\
&=& \langle (\partial^\gamma_l\partial^{\gamma,l}A^i - \partial^\gamma_l\partial^{\gamma,i}A^l)\nonumber\\
&& \times(\partial^\gamma_p\partial^{\gamma,p}A^j-\partial^\gamma_p\partial^{\gamma,j}A^p) \rangle.
\eeq
In momentum space,
\beq
C^{ij}(k) &=&  (k^2)^{2\gamma-2} \langle (k^2 A^i(k) - k_l k^iA^l(k)) \nonumber \\
&& \tabi \times(k^2 A^j(-k)-k_pk^jA^p(-k)) \rangle \nonumber \\
&=& (k^2)^{2\gamma-2} \bigg((k^2)^2 \langle A^i(k)A^j(-k)\rangle \nonumber \\
&& \tabi - k^2 k_pk^j \langle A^i(k)A^p(-k) \rangle \nonumber \\ 
&& \tabi - k^2 k_lk^i \langle A^l(k)A^j(-k)\rangle \nonumber \\
&& \tabi +  k_l k_p k^i k^j \langle A^l(k)A^p(-k) \rangle \bigg). \nonumber \\
\eeq
Using the propagator of $A^i$, we find
\beq
C^{ij}(k)  \propto (k^2)^{\gamma} \bigg( \eta^{ij} - \frac{k^ik^j}{k^2} \bigg).
\eeq
 Clearly, this equation not only obeys $k_\mu C^{\mu\nu}=0$ but also $k^{\gamma-1} k_\mu C^{\mu\nu}=0$. This translates into either $\partial_\mu C^{\mu\nu}=0$, the standard Ward identity, or
 \beq
 \partial_\mu (-\Delta)^{\frac{\gamma-1}{2}}C^{\mu\nu}=0
 \eeq
which illustrates beautifully the fact that the current conservation equation only specifies the current up to any operator that commutes with the total differential.  As we mentioned previously, this appears to be the first time this has been pointed out.  Consequently, nothing in this paper contradicts the standard linear conservation equations in electricity and magnetism nor in holography.  What is non-traditional is that the current now has an `anomalous'  (unorthodox) dimension.  In actuality it is more correct to refer 
to the dimension as unorthodox rather than anomalous because quantum corrections are irrelevant to the change in the dimension.

In a separate paper\cite{virasoroalg}, we have shown that currents possessing fractional dimensions based on the fractional gauge transformation of Eq. (\ref{eq:u1frac}) obey a multimodule Lie Virasoro algebra in which the generators, $L_n^a\equiv\left(\frac{\partial f}{\partial  z}\right)^a$, are governed by fractional derivatives of order $a\in {\mathbb R}$.  The Virasoro algebra is explicitly of the form,
\beq
[L^a_m,L_n^a]=A_{m,n}L^a_{m+n}+\delta_{m,n}h(n)cZ^a
\eeq
where $c$ is the central charge (not necessarily a constant), $Z^a$ is in the center of the algebra and $h(n)$ obeys a recursion relation related to the coefficients $A_{m,n}$.  As a result of this algebraic structure, currents based on the underlying gauge transformation, Eq. (\ref{eq:u1frac}), are described by a stable conformal IR fixed point.   From Eq. (\ref{fboundary}), it follows that the definition of the magnetic field
\begin{equation} \label{eq:fracA_fracB}
\vec\nabla^\alpha\times {_\alpha}\vec A= {_\alpha}\vec B
\end{equation}
involves the fractional curl. As a result, in simplifying the AB phase, 
\begin{equation}
\int {_\alpha}\vec B\cdot d \vec S \ne \oint {_\alpha}\vec A\cdot d \vec \ell,
\end{equation}
and as a consequence, the AB phase is no longer the traditional result. 

Theories with fractional gauge fields that preserve $U(1)$ are not uncommon.  Recently, one of us has shown\cite{GPWP} that they arise generically in bulk theories based on geodesically complete metrics anytime the bulk gauge field acquires a mass only along the holographic direction via the Higgs mechanism.  Consequently the boundary current complies with $U(1)$ invariance.  The resultant boundary theory contains fractional derivatives of the transverse components of the gauge field as proposed here.  The power of the derivative is determined by the mass and hence provides an additional length scale in the boundary theory.  Within the context of the renormalization group\cite{Nigel}, such a length scale is required for an anomalous dimension to exist. Hence, Ref. \cite{GPWP} provides a specific mechanism for realizing a theory with fractional gauge fields of the kind proposed here.   We also note that although the application of fractional calculus to the strange metal is new, numerous physical processes abound such as anomalous diffusion or Levy flights\cite{klafter} which have been described using fractional equations of motion.  We advocate here that the anomalies in the strange metal are tailor-made for fractional calculus.  

\section{Fractional Aharonov-Bohm Effect}

To derive the new result, we introduce a gauge connection into the Schr\"{o}dinger equation.  Let us define the covariant derivative $D_i \equiv \partial_i - i\frac{e}{\hbar}a_i$ with the associated gauge connection \cite{herrmann}
\begin{equation} \label{eq:def_little_a}
a_i \equiv [\partial_i,I^\alpha_i  {_\alpha} A_i] = \partial_i I^\alpha_i  {_\alpha}A_i
\end{equation}
where $I^\alpha$ is the fractional integral (see Appendices \ref{app:frac_calc_fourier} and \ref{app:frac_calc_real}).  The fundamental theorem of fractional calculus\cite{grigoletto} states that $I^\alpha \partial^\alpha\Lambda=\Lambda$.  As a consequence, 
 $a_\mu \rightarrow a_\mu+\partial_\mu\Lambda$ and our physical theory is gauge invariant although $a_\mu$ is directly related to the fractional gauge field.  Choosing $A_0=0$, we reduce the Schr\"{o}dinger equation to 
\begin{equation} \label{eq:schro_frac_A}
\bigg(-\frac{\hbar^2}{2m}(\partial_i - i\frac{e}{\hbar}a_i)^2+V\bigg)\psi = i\hbar\partial_t\psi.
\end{equation}
To derive the AB phase, let us consider a particle confined on the $x,y$ plane with a fractional magnetic field applied along the $z$ axis. Assume a particle can move from point $\vec r{_i}$ to $\vec r{_f}$ along path $\gamma_1$ (with wave function $\psi_1$) and along  path $\gamma_2$ (with wave function $\psi_2$). The total wave function at the point $\vec r{_f}$ at zero fractional magnetic field ($a_i = 0$) is $\psi = \psi_1 + \psi_2$. When the fractional magnetic field is turned on, the total wave function at $\vec r{_f}$ changes to
\begin{eqnarray}
\psi(\vec r_f,t) &=& e^{i\frac{e}{\hbar}\int_{\gamma_1}\vec a(\vec r)\cdot d \vec l}\psi_1(\vec r{_f},t) \nonumber \\
&& + e^{i\frac{e}{\hbar}\int_{\gamma_2}\vec a(\vec r)\cdot d \vec l}\psi_2(\vec r{_f},t) \nonumber\\
&=&  C \bigg( \psi_1(\vec r{_f},t) + e^{i\frac{e}{\hbar}\oint \vec a(\vec r)\cdot d \vec l} \psi_2(\vec r{_f},t) \bigg). \nonumber \\
\end{eqnarray}
Here $C$ is an over all phase factor $= e^{i\frac{e}{\hbar}\int_{\gamma_1} \vec a(\vec r)\cdot d \vec{l}}$. The phase difference between the two paths due to the gauge field is
\begin{equation} \label{eq:frac_AB_phase}
\Delta \phi = \frac{e}{\hbar}\oint \vec a(\vec r)\cdot d \vec l.
\end{equation}
In the strange metal, we posit that the current carrying degrees of freedom which emerge in the infrared couple to the fractional electromagnetic fields.  By definition, the propagating degrees of freedom are weakly interacting thereby warranting the Schr\"{o}dinger propagator approach we have adopted here. 
 
\begin{figure}[h!]
	\includegraphics[scale=0.5]{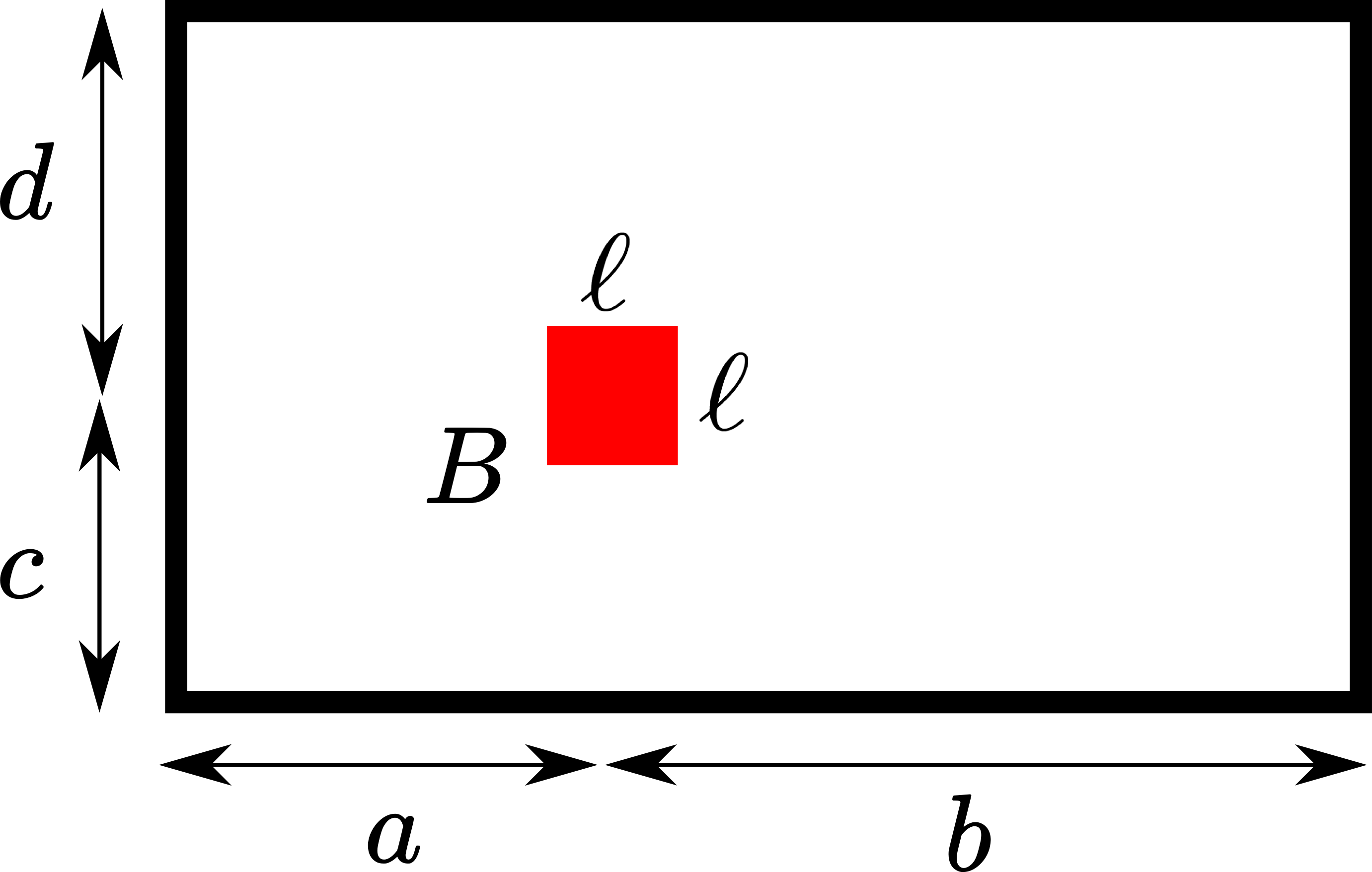} 
    \caption{Rectangle geometry that confines particle motion.  The fractional magnetic field is confined to the red region of size $\ell$ in the figure. } \label{fig:rectangle}
\end{figure}

\begin{figure}[h!] 
	\includegraphics[scale=0.5]{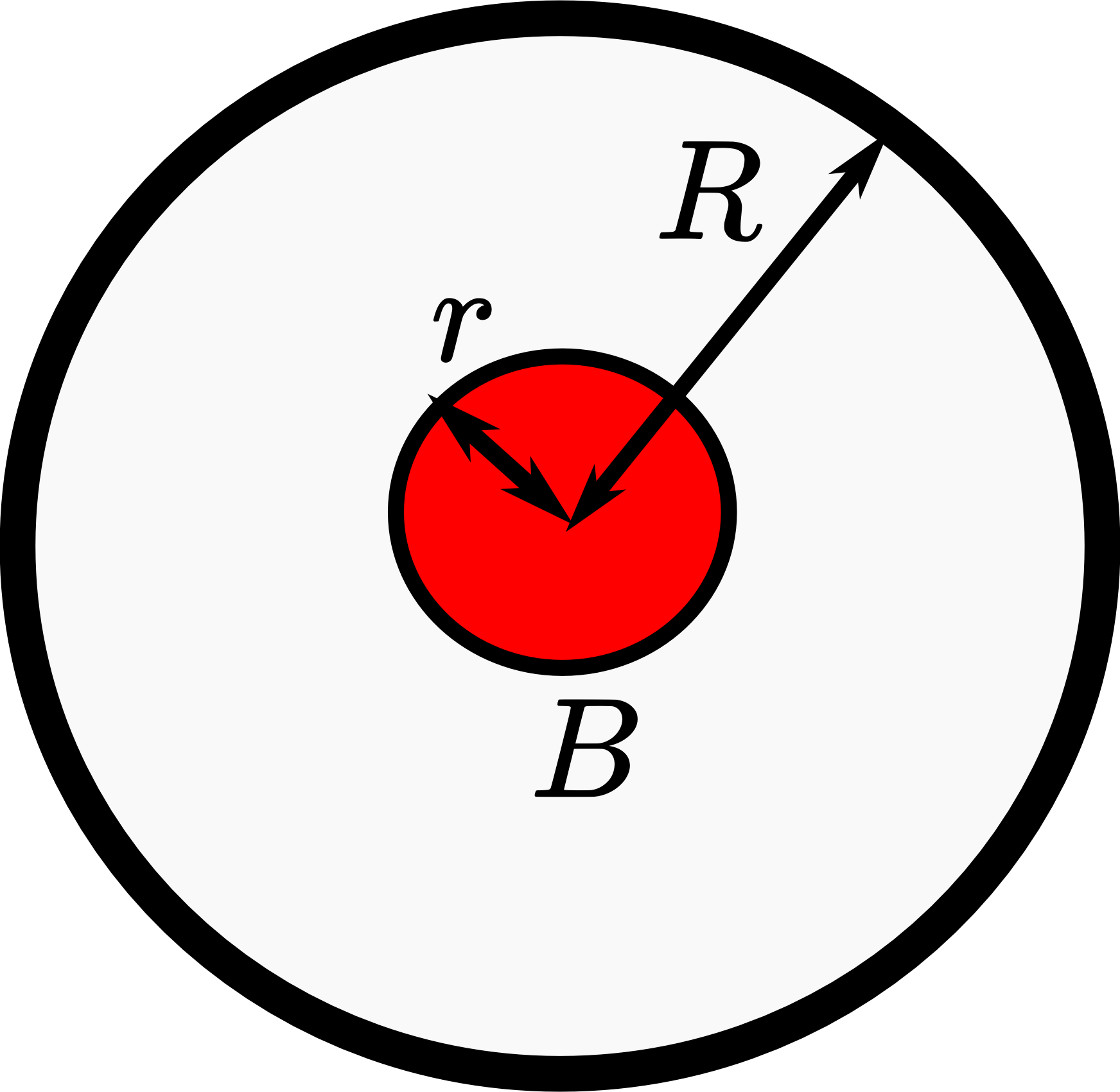} 
    \caption{Disk geometry for AB phase calculation.  The fractional magnetic field pierces the disk in a small region of radius, $r$. }  \label{fig:disk}
\end{figure}

We consider two different geometries in which an external magnetic field, $B$, pierces the sample and vanishes outside the shaded region in Figs. (\ref{fig:rectangle}) and (\ref{fig:disk}). We postulate that the $B$-field interacts with the material in such a way that the $B$-field acquires an anomalous dimension and hence becomes fractional, $_\alpha B$.\footnote{Depending on how $B$ and ${_\alpha}B$ are related in the material, there is a possibility of having finite magnetic monopoles in the system. However, it turns out that when $\alpha > 0$ magnetic monopoles does not exist for the field configurations we consider in Figs. (\ref{fig:rectangle}) and (\ref{fig:disk}). We discuss this issue in Appendix \ref{app:monopoles}} The charged particles in the sample now directly couple to the fractional vector potential $_\alpha \vec A$ instead of coupling to the external field $\vec A$. Hence, we can use Eq. (\ref{eq:frac_AB_phase}) to calculate the AB phase shift that these particles experience.

We work with five different definitions of fractional calculi (see Appendix \ref{app:frac_calc_fourier}). We show below only the result of the Feller calculus (for Fig. \ref{fig:rectangle}) and the rotationally invariant definition (for Fig. \ref{fig:disk}) because these definitions are odd under parity and thus the fractional gauge field formulated with these definitions will resemble the regular gauge field. The results for other definitions can be found in Appendices \ref{app:ab_rec} and \ref{app:ab_disk}. For the rectangle geometry in Fig. (\ref{fig:rectangle}), the AB phase for the Feller calculus when $a,b,c,d \gg \ell$ is
\beq \label{eq:phase_rect}
\Delta\phi_{R}= \frac{e}{\hbar}{_\alpha}B\ell^2 \bigg( \frac{(a^{\alpha-1}+b^{\alpha-1})(c^{\alpha-1}+d^{\alpha-1})}{4\Gamma^2(\alpha)\sin^2{\frac{\pi\alpha}{2}}} \bigg). \nonumber \\
\eeq
The phase picks up a geometric factor that is directly determined by the anomalous dimension $\alpha$ of the vector potential. The limiting value is $eB\ell^2/\hbar$ as $\alpha\rightarrow 1$. The convention that we have used is that the anomalous dimension is carried by the $_\alpha B$-field not the charge such that $[{_\alpha}B]=2\alpha$. As a result $\Delta\phi$ is dimensionless. 

The more experimentally tractable setup is most likely the disk in Fig. (\ref{fig:disk}). The AB phase shift for the rotationally invariant definition is:
\begin{widetext}  \begin{eqnarray} \label{eq:phase_circle}
\Delta \phi_{\rm D} = \frac{e}{\hbar}\pi r^2 {_\alpha}B R^{2\alpha-2} \left(\frac{2^{2-2\alpha}\Gamma(2-\alpha)}{\Gamma(\alpha)} {_2}F_1(1-\alpha,2-\alpha,2;\frac{r^2}{R^2})\right).
\end{eqnarray} \end{widetext}
Here  $_2F_1(a,b;c;z)$ is a hypergeometric function and the terms in the parenthesis reduce to unity in the limit $\alpha \rightarrow 1$. 

\section{Conclusion}

We have shown here that the presence of an anomalous dimension leads to a significant deviation from the standard AB phase. Appearing in the AB phase is a geometric factor in which the size of the sample is raised to a power involving the anomalous dimension. This extra sample-size dependence reflects the non-locality of the current. The correction is sizeable as it involves a ratio of the sample size to the region where the flux is threaded. As a result, we have provided an experimental diagnostic that is independent of any scaling ansatz.  One possible way to detect this AB phase is to perform a current interference experiment on a strange metal ring with a magnetic field at the center. This is the same geometry as Fig. \ref{fig:disk}. We predict that the periodicity of a magnetoresistance is directly proportional to the fractional AB phase as opposed to the standard AB phase. One can then extract the anomalous dimension by varying the ring's radius.  This setup is based on the experiment in which the standard AB phase was observed in a metallic ring\cite{Webb1985}.  Of course the success or failure of the experiment will be determined by how well phase coherence can be maintained along the excursion around the solenoid. Nonetheless, the clarity of our theoretical diagnostic should provide sufficient impetus for experiments along these lines to be performed which should serve to definitively settle that what is strange about the strange metal is that the current possesses an anomalous dimension.  

\noindent \textbf{Acknowledgements} P. Phillips thanks  Gabriele La Nave for numerous clarifications on gauge invariance and fractional calculus, J. Zaanen for comments on prior version of the manuscript, T. Senthil for a useful conversation, and the NSF DMR-1461952 for partial funding of this project and the Center for Emergent Superconductivity, a DOE Energy Frontier Research Center, Grant No. DE-AC0298CH1088. KL is supported by the Department of Physics at the University of Illinois and a scholarship from the Ministry of Science and Technology, Royal Thai Government. PP acknowledges support from the J. S. Guggenheim Foundation.

\onecolumngrid 

\appendix

\section{Fractional Calculus in Fourier-Space Representation}  \label{app:frac_calc_fourier}
Vector potentials with anomalous dimensions require fractional calculus. It can be defined by extending the standard integer derivatives and integrals to those involving fractional powers
\beq
\{I^n_x,\frac{\partial^n}{\partial x^n}\}\rightarrow \{I^\alpha_x,\frac{\partial^\alpha}{\partial x^\alpha}\}.
\eeq
Here $I^n_x$ is defined as a repeated integral $n$ times over $x$. We focus on five definitions of fractional calculi: left and right Liouville, Feller, Riesz\cite{grigoletto,Millercalc,herrmann_book,herrmann,Kilbas2006,Samko1993}, and the rotationally invariant definition. The rotationally invariant definition, $\partial^\alpha_i \equiv (-\nabla^2)^{\frac{\alpha-1}{2}}\partial_i$, is based on the fractional Laplacian and thus needs to be defined in dimensions greater or equal to two. These definitions of fractional calculi can be formulated in real or Fourier space.  For our purposes, it is most useful to implement the Fourier-space formulation,
\beq 
\partial_x^\alpha f(x) &=& \int\limits_{-\infty}^{\infty} \frac{dk}{2\pi} e^{ikx}F(\alpha,k)\tilde{f}(k) \label{eq:fourier_def_deriv}\\ 
I_x^\alpha f(x) &=& \int\limits_{-\infty}^{\infty} \frac{dk}{2\pi} e^{ikx}F^{-1}(\alpha,k)\tilde{f}(k) ,\label{eq:fourier_def_int}
\eeq
where $F(\alpha,k) = (ik)^\alpha$ for left Liouville, $F(\alpha,k) = (-ik)^\alpha$ for right Liouville, $F(\alpha,k) = i\mathrm{sgn}(k)|k|^\alpha$ for Feller, and $F(\alpha,k) = |k|^\alpha$ for Riesz. For the rotationally invariant definition of fractional calculus, one has the kernel for the fractional derivative/integral on the $x_i$ coordinate $F_i(\alpha,\vec k) = |\vec k|^{\alpha-1}ik_i$ with $\vec k$ being a $d$-dimensional momentum vector. Here $\partial_x^\alpha$ and $I_x^\alpha$ denote the fractional derivative and integral. The convention of the branch cut we use is $-\pi<\theta\le\pi$. Left and right Liouville are spatially asymmetric because, in real space, the operations involve an integration on the left and on the right of $x$, respectively (Eqs. (\ref{eq:left_deriv}) - (\ref{eq:right_int})). Feller calculus is odd under parity, and thus it resembles an odd-integer-order calculus.  On the other hand, since Riesz calculus is even under parity, its behavior is similar to an even-integer-order calculus. The rotationally invariant definition is rotational invariant and odd under parity. In terms of formal mathematical operations, the methods outlined have restrictions regarding the range of validity of $\alpha$. For both left and right Liouville calculi, one needs $0<\alpha<1$. For the Feller and the Reisz calculi, one needs $0<\alpha<2$. Nonetheless, the results can be analytically continued outside this range. 

The important property of these definition is that when $\alpha>0$ the fractional derivative of a constant is zero. Let $f(x) = C$ where $C$ is a constant. The Fourier component of $f(x)$ is $\tilde{f}(k) = 2\pi C\delta(k)$. Consequently, 
\beq \label{eq:frac_deriv_const}
\partial_x^\alpha f(x) = CF(\alpha,0) = 0.
\eeq
For other definitions such as the left Riemann derivative (Eq. (\ref{eq:real_lrl_deriv}) with $a = 0$) and the right Riemann derivative (Eq. (\ref{eq:real_rrl_deriv}) with $b = 0$), $\partial_x^\alpha f(x)$ can be nonzero. 

\section{Fractional Calculus in Coordinate-Space Representation} \label{app:frac_calc_real}

The fractional calculi in the previous section are formulated in the Fourier-space representations. Alternatively, they can be defined in coordinate space \cite{grigoletto,Millercalc,herrmann_book,herrmann,Kilbas2006,Samko1993}. Let $a$ and $b$ be real numbers. We define the following notations for the fractional derivative and the corresponding integral for $x>a$:
\beq 
D^x_a(\alpha) f(x) &=& \frac{1}{\Gamma(n-\alpha)}\frac{d^n}{dx^n}\int\limits_{a}^{x}dx' (x-x')^{n-\alpha-1}f(x') \label{eq:left_deriv}\\
I^x_a(\alpha) f(x) &=& \frac{1}{\Gamma(\alpha)}\int\limits_{a}^{x} dx'(x-x')^{\alpha-1} f(x'). \label{eq:left_int}
\eeq
When $x < b$, we define 
\beq
D^b_x(\alpha) f(x) &=& \frac{1}{\Gamma(n-\alpha)}\bigg(-\frac{d}{dx}\bigg)^n\int\limits_{x}^{b}dx' (x'-x)^{n-\alpha-1}f(x') \label{eq:right_deriv}\\
I^b_x(\alpha) f(x) &=& \frac{1}{\Gamma(\alpha)}\int\limits_{x}^{b} dx'(x'-x)^{\alpha-1} f(x') \label{eq:right_int}
\eeq
where $n = [\alpha]+1$ and $[\alpha]$ denotes the integer part of $\alpha$.

The left Riemann-Liouville fractional calculus corresponds to
\beq
D^\alpha_{LRL} &=& D^x_{a}(\alpha) \label{eq:real_lrl_deriv} \\
I^\alpha_{LRL} &=& I^x_{a}(\alpha) \label{eq:real_lrl_int},
\eeq
while the right Riemann-Liouville fractional calculus is
\beq 
D^\alpha_{RRL} &=& D^{b}_x(\alpha) \label{eq:real_rrl_deriv}\\
I^\alpha_{RRL} &=& I^{b}_x(\alpha). \label{eq:real_rrl_int}
\eeq
{\it The Liouville fractional calculi} is the special case of the Riemann-Liouville calculi with $a = -\infty$ and $b = \infty$.

The Feller fractional calculus corresponds to
\beq
D^\alpha_{F} &=& \frac{1}{2\sin\frac{\pi\alpha}{2}}(D^x_{-\infty}(\alpha)-D_x^{\infty}(\alpha)) \label{eq:real_feller_deriv} \\
I^\alpha_{F} &=& \frac{1}{2\sin\frac{\pi\alpha}{2}}(I^x_{-\infty}(\alpha)-I_x^{\infty}(\alpha)) \label{eq:real_feller_int}
\eeq
and the Riesz fractional calculus corresponds to
\beq 
D^\alpha_{RZ} &=& \frac{1}{2\cos\frac{\pi\alpha}{2}}(D^x_{-\infty}(\alpha)+D_x^{\infty}(\alpha)) \label{eq:real_riesz_deriv} \\
I^\alpha_{RZ} &=& \frac{1}{2\cos\frac{\pi\alpha}{2}}(I^x_{-\infty}(\alpha)+I_x^{\infty}(\alpha)). \label{eq:real_riesz_int}
\eeq

The Fourier-space formulations can be shown to be the same as the coordinate space representation. We explicitly show this for the case of the left Liouville calculus. We start by rewriting Eq. (\ref{eq:fourier_def_int}) in the case of left Liouville to
\beq
I^\alpha_{\mathrm{LL}} f(x) &=& \int\limits_{-\infty}^{\infty}dx' K(x-x')f(x')
\eeq
where the kernel $K(x-x') = \int\limits_{-\infty}^{\infty}\frac{dk}{2\pi}e^{ik(x-x')}(ik)^{-\alpha}$ and the subscript LL denotes left Liouville. This integral can be evaluated to be
\beq
K(x-x') = \Theta(x-x')\frac{(x-x')^{\alpha-1}}{\Gamma(\alpha)}
\eeq
when $0 < \alpha < 1$. Thus, the left Liouville integral in coordinate space is 
\beq \label{eq:LL}
I^\alpha_{\mathrm{LL}} f(x) = \frac{1}{\Gamma(\alpha)}\int\limits_{-\infty}^{x}dx'(x-x')^{\alpha-1}f(x') = I^x_{-\infty}(\alpha).
\eeq
Similarly, we rewrite the left Liouville derivative from Eq. (\ref{eq:fourier_def_deriv}) to
\beq
\partial^\alpha_{\mathrm{LL}} f(x) &=& \int\limits_{-\infty}^{\infty} \frac{dk}{2\pi} e^{ikx}(ik)^{\alpha}\tilde{f}(k) \nonumber \\
&=& \frac{d^n}{dx^n}\int\limits_{-\infty}^{\infty} \frac{dk}{2\pi} e^{ikx}(ik)^{-(n-\alpha)}\tilde{f}(k) \nonumber \\
&=& \frac{d^n}{dx^n}I^{n-\alpha}_{\mathrm{LL}} f(x) \nonumber \\
&=& \frac{1}{\Gamma(n-\alpha)}\frac{d^n}{dx^n}\int\limits_{-\infty}^{x}dx'(x-x')^{n-\alpha-1}f(x') \nonumber \\
&=& D^x_{-\infty}(\alpha)f(x)
\eeq
where $n = [\alpha] + 1$. The equivalences between the Fourier-space and the coordinate-space formulations of the right Liouville, Feller, and Riesz can be shown in similar manner. 

\section{Absence of Magnetic Monopoles in Constant Fractional Magnetic Field} \label{app:monopoles}

Let us consider the possibility of having magnetic monopoles (or magnetic charges) in a fractional electromagnetic system.\footnote{We mean here the system in which its gauge transformation is defined according to Eq. (\ref{eq:u1frac}).} Let ${_\alpha}\vec B$ denote the fractional magnetic field and $\vec B$ denotes the actual magnetic field. One can define the magnetic charge density $\rho_m$ as a fractional divergence of the fractional magnetic field,
\beq
\rho_m = \vec\nabla^\alpha\cdot{_\alpha}\vec B.
\eeq
The question whether $\rho_m$ equals zero depends on how one associates ${_\alpha}\vec B$ with $\vec B$ and on the definition of the fractional derivative we consider. We focus on the four definitions discussed in Appendix \ref{app:frac_calc_fourier}. If we assume ${_\alpha}\vec B \propto \vec B$, then $\vec\nabla\cdot {_\alpha} \vec B = 0$. However, this does not necessarily imply that $\vec\nabla^\alpha\cdot {_\alpha}\vec B = 0$. So it is possible to have a nonzero $\rho_m$. 

It turns out that for the field configurations in Figs. (\ref{fig:rectangle}) and (\ref{fig:disk}) $\rho_m$ vanishes when $\alpha>0$. From Eqs. (\ref{eq:frac_mag_rect}) and (\ref{eq:frac_mag_disk}), we have
\beq
{_\alpha}\vec{B}(x,y,z) = {_\alpha}B_z\hat{z},
\eeq
with ${_\alpha}B_z =  {_\alpha}B\Theta(\ell^2/4-x^2)\Theta(\ell^2/4-y^2)$ for the rectangle geometry and ${_\alpha}B_z = {_\alpha}B\Theta(r-\sqrt{x^2+y^2})$ for the disk geometry.
We can directly compute $\rho_m$ by taking the fractional divergence. We find that
\beq
\rho_m = \vec\nabla^\alpha\cdot {_\alpha}\vec B = \partial^\alpha_z {_\alpha}B_z = 0,
\eeq
with $\alpha > 0$ and we have used Eq. (\ref{eq:frac_deriv_const}) since $B_z$ does not depend on $z$.  Consequently, for the system considered here magnetic monopoles do not exist.

\section{Fractional Aharonov-Bohm Effect in Rectangular geometry} \label{app:ab_rec}

The expression for ${_\alpha}\vec{B}$ from Fig. (\ref{fig:rectangle}) is
\beq
{_\alpha}\vec{B}(x,y) = {_\alpha}B\Theta(\ell^2/4-x^2)\Theta(\ell^2/4-y^2)\hat{z}. \label{eq:frac_mag_rect}
\eeq
The Fourier transform of ${_\alpha}\vec{B}(x,y)$ is
\beq
{_\alpha}\vec{B}(\vec{k}) = {_\alpha}B_z(\vec{k})\hat{z} = 4({_\alpha}B)\frac{\sin{\frac{k_x\ell}{2}}\sin{\frac{k_y\ell}{2}}}{k_xk_y}\hat{z}.
\eeq
Below we directly use the Fourier-space formulations to evaluate fractional derivatives and integrals.

\subsection{Left Liouville Fractional Calculus}
We solve ${_\alpha}\vec{A}(\vec{k})$ from
\beq \label{eq:B_curlA}
{_\alpha}\vec{B}(\vec{k}) = (i\vec{k})^{\alpha} \times {_\alpha}\vec{A}(\vec{k})
\eeq
where $(i\vec{k})^{\alpha} = \{(ik_x)^\alpha,(ik_y)^\alpha,0 \}$. A choice of ${_\alpha}\vec{A}(\vec{k})$ that satisfies Eq. (\ref{eq:B_curlA}) is 
\beq
{_\alpha}\vec{A}(\vec{k}) = \frac{{_\alpha}B_z(\vec{k})}{(ik_x)^{2\alpha}+(ik_y)^{2\alpha}}\{-(ik_y)^\alpha, (ik_x)^\alpha, 0 \}.
\eeq
Next, using Eq. (7) of the main text, we obtain $\vec{a}(\vec{k})$ as
\beq
\vec{a}(\vec{k}) = \frac{{_\alpha}B_z(\vec{k})}{(ik_x)^{2\alpha}+(ik_y)^{2\alpha}}\{-(ik_x)^{1-\alpha}(ik_y)^\alpha, (ik_x)^\alpha(ik_y)^{1-\alpha}, 0 \}. 
\eeq
It is easiest to work with $\vec{b}(\vec{k}) = (i\vec{k}) \times \vec{a}(\vec{k}) $.  We obtain
\beq
\vec{b}(\vec{k}) &=& {_\alpha}B_z(\vec{k})(ik_x)^{1-\alpha}(ik_y)^{1-\alpha} \hat{z} \nonumber \\
 &=& -4({_\alpha}B)\sin{\frac{k_x\ell}{2}}\sin{\frac{k_y\ell}{2}}(ik_x)^{-\alpha}(ik_y)^{-\alpha} \hat{z},
\eeq
which in  position space becomes
\begin{eqnarray} \label{eq:b_z_LL}
b_z(x,y) &=& \int\limits_{-\infty}^{\infty}\frac{dk_x}{2\pi}\frac{dk_y}{2\pi}b_z(\vec{k}) \nonumber \\
&=& 4({_\alpha}B)\ell^{2\alpha-2}f_1(\frac{x}{\ell})f_1(\frac{y}{\ell}),
\end{eqnarray}
where 
\beq
f_1(s) &=& \int\limits_{-\infty}^{\infty} \frac{dz}{2\pi}i(iz)^{-\alpha}\sin\frac{z}{2}e^{izs}\nonumber\\
&=& \frac{1}{2\Gamma(\alpha)}\bigg( \Theta(s+\frac{1}{2})(s+\frac{1}{2})^{\alpha-1} - \Theta(s-\frac{1}{2})(s-\frac{1}{2})^{\alpha-1}\bigg).
\eeq
Consequently, we obtain
\beq
b_z(x,y) = && \frac{{_\alpha}B}{\Gamma^2(\alpha)}\bigg( \Theta(x+\frac{\ell}{2})(x+\frac{\ell}{2})^{\alpha-1} - \Theta(x-\frac{\ell}{2})(x-\frac{\ell}{2})^{\alpha-1}\bigg) \nonumber \\
&& \times \bigg( \Theta(y+\frac{\ell}{2})(y+\frac{\ell}{2})^{\alpha-1} - \Theta(y-\frac{\ell}{2})(y-\frac{\ell}{2})^{\alpha-1}\bigg).
\eeq
The phase difference is
\begin{eqnarray}
\Delta \phi &=& \frac{e}{\hbar} \int\limits_{-a}^{b}dx\int\limits_{-c}^{d}dyb_z(x,y) \nonumber \\
&=& \frac{e}{\hbar\alpha^2\Gamma^2(\alpha)}{_\alpha}Bb^{\alpha}d^{\alpha}\bigg((1+\frac{\ell}{2b})^{\alpha} - (1-\frac{\ell}{2b})^{\alpha} \bigg)\bigg((1+\frac{\ell}{2d})^{\alpha} - (1-\frac{\ell}{2d})^{\alpha} \bigg).
\end{eqnarray}
In the limit $b\gg \ell$ and $d \gg \ell$, 
\begin{eqnarray}
\Delta \phi &\approx& \frac{e}{\hbar}{_\alpha}B\ell^2\bigg(\frac{b^{\alpha-1}d^{\alpha-1}}{\Gamma^2(\alpha)} \bigg).
\end{eqnarray}
The AB phase from the left Liouville calculus is not symmetric. It involves only the length $b$ and $d$, but not $a$ and $c$. This result can be understood from the fact that the left Liouville calculus is spatially asymmetric.

\subsection{Right Liouville Fractional Calculus}
The resulting $b_z(\vec{k})$ is the same as Eq. (\ref{eq:b_z_LL}) but the function $f_1(s)$ is replaced with 
\beq
f_2(s) &=& \int\limits_{-\infty}^{\infty} \frac{dz}{2\pi}i(-iz)^{-\alpha}\sin\frac{z}{2}e^{izs}\nonumber\\
& =& \frac{1}{2\Gamma(\alpha)}\bigg( \Theta(-s-\frac{1}{2})(-s-\frac{1}{2})^{\alpha-1} - \Theta(-s+\frac{1}{2})(-s+\frac{1}{2})^{\alpha-1}\bigg).
\eeq
Performing the area integral, we find that
\beq
\Delta \phi &=& \frac{e}{\hbar\alpha^2\Gamma^2(\alpha)}{_\alpha}Ba^{\alpha}c^{\alpha}\bigg((1+\frac{\ell}{2a})^{\alpha} - (1-\frac{\ell}{2a})^{\alpha} \bigg)\bigg((1+\frac{\ell}{2c})^{\alpha} - (1-\frac{\ell}{2c})^{\alpha} \bigg).
\eeq
In the limit of $a\gg \ell$ and $c \gg \ell$,
\beq
\Delta \phi &\approx& \frac{e}{\hbar}{_\alpha}B\ell^2\bigg(\frac{a^{\alpha-1}c^{\alpha-1}}{\Gamma^2(\alpha)} \bigg).
\eeq
As in the case of the Left Liouville calculus, the phase is not symmetric, because the right Liouville calculus is also spatially asymmetric.

\subsection{Feller Fractional Calculus}
The resulting $b_z(\vec{k})$ is the same as Eq. (\ref{eq:b_z_LL}) but the function $f_1(s)$ is replaced with 
\beq
f_3(s) &= &\int\limits_{-\infty}^{\infty}\frac{dz}{2\pi}\mathrm{sgn}(z)|z|^{-\alpha}\sin{\frac{z}{2}}e^{izs}\nonumber\\
&=& -\frac{1}{4\Gamma(\alpha)\sin{\frac{\pi\alpha}{2}}}\bigg(\Theta(s+\frac{1}{2})(s+\frac{1}{2})^{\alpha-1} - \Theta(-s-\frac{1}{2})(-s-\frac{1}{2})^{\alpha-1} \nonumber \\ 
&& \tabi \ \ \ \ \ \ \ \ - \Theta(s-\frac{1}{2})(s-\frac{1}{2})^{\alpha-1}  + \Theta(-s+\frac{1}{2})(-s+\frac{1}{2})^{\alpha-1} \bigg) .
\eeq
The phase difference is
\begin{eqnarray}
\Delta\phi &=& \frac{e({_\alpha}B)}{4\hbar\alpha^2\Gamma^2(\alpha)\sin^2{\frac{\pi\alpha}{2}}} \bigg( a^\alpha[(1+\frac{\ell}{2a})^\alpha - (1-\frac{\ell}{2a})^\alpha] + b^\alpha[(1+\frac{\ell}{2b})^\alpha - (1-\frac{\ell}{2b})^\alpha] \bigg) \nonumber \\
&& \times \bigg( c^\alpha[(1+\frac{\ell}{2c})^\alpha - (1-\frac{\ell}{2c})^\alpha] + d^\alpha[(1+\frac{\ell}{2d})^\alpha - (1-\frac{\ell}{2d})^\alpha] \bigg),
\end{eqnarray}
which in the limit of $a,b,c,d \gg \ell$, reduces to
\beq
\Delta \phi \approx \frac{e({_\alpha}B)\ell^2}{\hbar} \bigg( \frac{(a^{\alpha-1}+b^{\alpha-1})(c^{\alpha-1}+d^{\alpha-1})}{4\Gamma^2(\alpha)\sin^2{\frac{\pi\alpha}{2}}} \bigg).
\eeq

\subsection{Riesz Fracational Calculus}
The resulting $b_z(\vec{k})$ is the same as Eq. (\ref{eq:b_z_LL}) but the function $f_1(s)$ is replaced with 
\beq
f_4(s)& =& \int\limits_{-\infty}^{\infty}\frac{dz}{2\pi}i|z|^{-\alpha}\sin{\frac{z}{2}}e^{izs}\nonumber\\
&=& \frac{1}{4\Gamma(\alpha)\cos{\frac{\pi\alpha}{2}}}\bigg(\Theta(s+\frac{1}{2})(s+\frac{1}{2})^{\alpha-1} + \Theta(-s-\frac{1}{2})(-s-\frac{1}{2})^{\alpha-1} \nonumber \\
&& \tabi \ \ \ \ \ \ \ \ - \Theta(s-\frac{1}{2})(s-\frac{1}{2})^{\alpha-1} - \Theta(-s+\frac{1}{2})(-s+\frac{1}{2})^{\alpha-1} \bigg). 
\eeq
The phase difference is
\begin{eqnarray}
\Delta\phi &=& \frac{e}{4\hbar\alpha^2\Gamma^2(\alpha)\cos^2{\frac{\pi\alpha}{2}}}{_\alpha}B\bigg( a^\alpha[(1+\frac{\ell}{2a})^\alpha - (1-\frac{\ell}{2a})^\alpha] - b^\alpha[(1+\frac{\ell}{2b})^\alpha - (1-\frac{\ell}{2b})^\alpha] \bigg) \nonumber \\
&& \times \bigg( c^\alpha[(1+\frac{\ell}{2c})^\alpha - (1-\frac{l}{2c})^\alpha] - d^\alpha[(1+\frac{\ell}{2d})^\alpha - (1-\frac{\ell}{2d})^\alpha] \bigg).
\end{eqnarray}
In the limit of $a,b,c,d \gg \ell$,
\beq \label{eq:phase_rect_riesz}
\Delta \phi \approx \frac{e}{\hbar}{_\alpha}B\ell^2 \bigg( \frac{(a^{\alpha-1}-b^{\alpha-1})(c^{\alpha-1}-d^{\alpha-1})}{4\Gamma^2(\alpha)\cos^2{\frac{\pi\alpha}{2}}} \bigg).
\eeq
The limiting value of the phase is not $eB\ell^2/\hbar$ as $\alpha\rightarrow 1$. We can understand this result from the fact that the Riesz calculus has an even parity, so one cannot expect it to have the same behavior as the first order derivative.

\section{Fractional Aharonov-Bohm Effect of Disk Geometry} \label{app:ab_disk}

We consider now the disk geometry shown in Fig. (\ref{fig:disk}).  The fractional magnetic field is given by
\beq
{_\alpha}\vec{B} = {_\alpha}B_z(\vec{\rho})\hat{z} = {_\alpha}B\Theta(r-\rho)\hat{z}.  \label{eq:frac_mag_disk}
\eeq
In Fourier space,
\beq
{_\alpha}B_z(\vec{k}) = {_\alpha}B\int\limits_{-\infty}^{\infty}dx \int\limits_{-\infty}^{\infty}dy e^{-i\vec{k} \cdot \vec{\rho}}\Theta(r-\rho).
\eeq
We now change to polar coordinates, $\vec{\rho} = \rho\cos\phi \hat{x} + \rho\sin\phi\hat{y}$ and $\vec{k} = k\cos\xi \hat{x} + k\sin\xi\hat{y}$. The result is
\begin{eqnarray}
{_\alpha}B_z(\vec{k}) &=& {_\alpha}B\int\limits_{0}^{2\pi}d\phi \int\limits_{0}^{r}d\rho \rho e^{-ik\rho\cos(\phi-\xi)} \nonumber \\
&=& \frac{2\pi r}{k}{_\alpha}BJ_1(kr)
\end{eqnarray}

\subsection{Left Liouville Fractional Calculus} \label{app:disk_left}
We perform the same calculation as in the rectangle case to obtain
\begin{eqnarray}
b_z(\vec{k}) &=& {_\alpha}B_z(\vec{k})(ik_x)^{1-\alpha}(ik_y)^{1-\alpha} \nonumber  \\
&=& 2\pi r{_\alpha}Bk^{1-2\alpha}J_1(kr)(i\cos\xi)^{1-\alpha}(i\sin\xi)^{1-\alpha}.
\end{eqnarray}
In position space,
\beq
b_z(\rho,\theta) = \frac{1}{4\pi^2}\int\limits_{0}^{\infty}dk \int\limits_{0}^{2\pi}d\xi e^{ik\rho\cos{(\theta-\xi)}}k^{2-2\alpha}2\pi r{_\alpha}BJ_1(kr)(i\cos\xi)^{1-\alpha}(i\sin\xi)^{1-\alpha}.
\eeq
The phase difference is the area integral of $b_z$ over the disk of radius $R$ in Fig. 2,
\begin{eqnarray}
\Delta\phi &=& \frac{e}{\hbar}\int\limits_{0}^{R}d\rho\int\limits_{0}^{2\pi}d\theta \rho b_z(\rho,\theta) \nonumber \\
&=&  \frac{er{_\alpha}B}{2\pi\hbar}\int\limits_{0}^{\infty}dk\int\limits_{0}^{2\pi}d\xi\int\limits_{0}^{R}d\rho\int\limits_{0}^{2\pi}d\theta \rho  k^{2-2\alpha} e^{ik\rho\cos(\theta-\xi)}J_1(kr) (i\cos\xi)^{1-\alpha}(i\sin\xi)^{1-\alpha}.
\end{eqnarray}
The $\theta$ integration yields  
\beq
\int\limits_{0}^{2\pi}d\theta e^{ik\rho\cos(\theta-\xi)} = 2\pi J_0(k\rho). 
\eeq
Consequently, we reduce the phase difference to
\begin{eqnarray}
\Delta\phi = \frac{er{_\alpha}B}{\hbar}\int\limits_{0}^{\infty}dk\int\limits_{0}^{2\pi}d\xi\int\limits_{0}^{R}d\rho \rho  k^{2-2\alpha} J_1(kr)J_0(k\rho)(i\cos\xi)^{1-\alpha}(i\sin\xi)^{1-\alpha}.
\end{eqnarray}
The integral over $\rho$ can be done analytically,
\beq
\int\limits_{0}^{R}d\rho \rho J_0(k\rho) = \frac{R}{k}J_1(kR).
\eeq
Hence, the phase difference becomes
\begin{eqnarray}
\Delta\phi = \frac{er{_\alpha}BR}{\hbar}\int\limits_{0}^{\infty}dkk^{1-2\alpha} J_1(kr)J_1(kR) \int\limits_{0}^{2\pi}d\xi (i\cos\xi)^{1-\alpha}(i\sin\xi)^{1-\alpha}.
\end{eqnarray}
The two integrals can be evaluated as
\beq
\int\limits_{0}^{\infty}dkk^{1-2\alpha} J_1(kr)J_1(kR) = \frac{2^{1-2\alpha}rR^{2\alpha-3}\Gamma(2-\alpha)}{\Gamma(\alpha)}{_2}F_1(1-\alpha,2-\alpha;2;(\frac{r}{R})^2)
\eeq
and
\beq
\int\limits_{0}^{2\pi}d\xi (i\cos\xi)^{1-\alpha}(i\sin\xi)^{1-\alpha} = \frac{2^\alpha\sin^2\frac{\pi\alpha}{2}\sqrt{\pi}\Gamma(1-\frac{\alpha}{2})}{\Gamma(\frac{3}{2}-\frac{\alpha}{2})}.
\eeq
Here $_2F_1(a,b;c;z)$ is a hypergeometric function. Finally, the phase difference is
\beq \label{eq:phase_circle_left}
\Delta\phi = \frac{e}{\hbar}\pi r^2 {_\alpha}B R^{2\alpha-2}\Bigg(\frac{2^{1-\alpha}\Gamma(2-\alpha)\Gamma(1-\frac{\alpha}{2})}{\sqrt{\pi}\Gamma(\alpha)\Gamma(\frac{3}{2}-\frac{\alpha}{2})}\sin^2\frac{\pi\alpha}{2}{_2}F_1(1-\alpha,2-\alpha;2;\frac{r^2}{R^2}) \Bigg)
\eeq
The terms in the parenthesis reduce to 1 in the limit $\alpha \rightarrow 1$. 

\subsection{Right Liouville Fractional Calculus}
The phase difference from this fractional calculus is the same as the phase in Eq. (\ref{eq:phase_circle_left}) because one can show that
\beq
b_z(\vec{k}) = 2\pi r{_\alpha}Bk^{1-2\alpha}J_1(kr)(-i\cos\xi)^{1-\alpha}(-i\sin\xi)^{1-\alpha}
\eeq
and the integral
\beq
\int\limits_{0}^{2\pi}d\xi (-i\cos\xi)^{1-\alpha}(-i\sin\xi)^{1-\alpha} = \frac{2^\alpha\sin^2\frac{\pi\alpha}{2}\sqrt{\pi}\Gamma(1-\frac{\alpha}{2})}{\Gamma(\frac{3}{2}-\frac{\alpha}{2})}.
\eeq

\subsection{Feller Fractional Calculus}
For this definition, one can show that
\beq
b_z(\vec{k}) = 2\pi r{_\alpha}Bk^{1-2\alpha}J_1(kr)|\cos\xi|^{1-\alpha}|\sin\xi|^{1-\alpha}.
\eeq
The only difference from the right Liouville calculus is the integration over $\xi$. One finds
\beq
\int\limits_{0}^{2\pi}d\xi |\cos\xi|^{1-\alpha}|\sin\xi|^{1-\alpha} = \frac{2^\alpha\sqrt{\pi}\Gamma(1-\frac{\alpha}{2})}{\Gamma(\frac{3}{2}-\frac{\alpha}{2})}.
\eeq
And hence the phase difference is
\beq
\Delta\phi = \frac{e}{\hbar}\pi r^2 {_\alpha}B R^{2\alpha-2}\Bigg(\frac{2^{1-\alpha}\Gamma(2-\alpha)\Gamma(1-\frac{\alpha}{2})}{\sqrt{\pi}\Gamma(\alpha)\Gamma(\frac{3}{2}-\frac{\alpha}{2})}{_2}F_1(1-\alpha,2-\alpha;2;\frac{r^2}{R^2}) \Bigg).
\eeq

\subsection{Riesz Fractional Calculus}
For this definition, one can show that
\beq
b_z(\vec{k}) = -2\pi r{_\alpha}Bk^{1-2\alpha}J_1(kr)\cos\xi|\cos\xi|^{-\alpha}\sin\xi|\sin\xi|^{-\alpha}.
\eeq
The integral over $\xi$ vanishes because $\cos\xi|\cos\xi|^{-\alpha}\sin\xi|\sin\xi|^{-\alpha}$ is an odd function. As a result
\beq
\Delta\phi = 0.
\eeq
This result is not surprising, because from Eq. (\ref{eq:phase_rect_riesz}), the AB phase from the Riesz calculus when $a = b$ and $c = d$ is zero. 

\subsection{Rotationally Invariance Definition}
The fractional Laplacian in the definition, $\partial^\alpha_i = (-\nabla^2)^{\frac{\alpha-1}{2}}\partial_i$, is to be interpreted as a two-dimensional operator. Hence, in the kernel $F_i(\alpha,\vec k) = |\vec k|^{\alpha-1}ik_i$, one has $|\vec k|^2 = k_x^2+k_y^2$. The calculation is proceeded in the same manner as what we have done for other definitions. One can show that
\beq
b_z(\vec k) = k^{1-2\alpha}2\pi r {_\alpha}B J_1(kr).
\eeq
Unlike other definitions, there is no dependence on $\xi$ because this definition is rotationally invariance. The phase shift is
\beq
\Delta \phi = \frac{e}{\hbar}\pi r^2 {_\alpha}B R^{2\alpha-2} \left(\frac{2^{2-2\alpha}\Gamma(2-\alpha)}{\Gamma(\alpha)} {_2}F_1(1-\alpha,2-\alpha,2;\frac{r^2}{R^2})\right).
\eeq

\twocolumngrid

%\bibliography{anombibfinal0}
%\bibliographystyle{apsrev4-1}
%merlin.mbs apsrev4-1.bst 2010-07-25 4.21a (PWD, AO, DPC) hacked
%Control: key (0)
%Control: author (72) initials jnrlst
%Control: editor formatted (1) identically to author
%Control: production of article title (-1) disabled
%Control: page (0) single
%Control: year (1) truncated
%Control: production of eprint (0) enabled
%
\end{document}